\documentclass[aps,prl,superscriptaddress,showpacs,twocolumn]{revtex4-1}
\usepackage[pdftex]{graphicx}
\usepackage{amsmath}
\usepackage{xspace}
\usepackage{color}
\usepackage[colorlinks=true,linkcolor=blue,urlcolor=blue,citecolor=blue]{hyperref}
\usepackage[normalem]{ulem}
\usepackage{hyperref}

\newcommand{\ie}{i.e.\@\xspace}

\newcommand{\eqw}[1]{(\ref{#1})}
\newcommand{\eq}[1]{Eq.\thinspace{}(\ref{#1})}

\newcommand{\fig}[1]{Fig.\thinspace{}\ref{#1}}

\newcommand{\fc}[1]{({#1})}
\newcommand{\figc}[2]{Fig.\thinspace{}\ref{#1}\thinspace{}\fc{#2}}

\usepackage{color}
\definecolor{BV}{rgb}{0.1,0.,0.6}
\definecolor{R}{rgb}{0.9,0,0}
\definecolor{G}{rgb}{0.2,0.8,0.2}
\usepackage{comment}

\begin{document}

\title{Quantum Flutter: Signatures and Robustness}

\author{Michael Knap}
\affiliation{Department of Physics, Harvard University, Cambridge, MA 02138, USA}
\affiliation{ITAMP, Harvard-Smithsonian Center for Astrophysics, Cambridge, MA 02138, USA}
\affiliation{Institute of Theoretical and Computational Physics, Graz University of Technology, 8010 Graz, Austria}

\author{Charles J. M. Mathy}
\affiliation{ITAMP, Harvard-Smithsonian Center for Astrophysics, Cambridge, MA 02138, USA}
\affiliation{Department of Physics, Harvard University, Cambridge, MA 02138, USA}

\author{Martin Ganahl}
\affiliation{Institute of Theoretical and Computational Physics, Graz University of Technology, 8010 Graz, Austria}

\author{Mikhail B. Zvonarev}
\affiliation{Univ Paris-Sud, Laboratoire LPTMS, UMR8626, Orsay, F-91405, France}
\affiliation{CNRS, Orsay, F-91405, France}

\author{Eugene Demler}
\affiliation{Department of Physics, Harvard University, Cambridge, MA 02138, USA}

\date{\today}

\begin{abstract}

We investigate the motion of an impurity particle injected with finite velocity into an 
interacting one-dimensional quantum gas. Using large-scale numerical simulations based 
on matrix product states, we observe and quantitatively analyze long-lived oscillations 
of the impurity momentum around a non-zero saturation value, called quantum flutter. We 
show that the quantum flutter frequency is equal to the energy difference between two 
branches of collective excitations of the model. We propose an explanation of the finite 
saturation momentum of the impurity based on the properties of the edge of the excitation 
spectrum. Our results indicate that quantum flutter exists away from integrability, and 
provide parameter regions in which it could be observed in experiments with ultracold 
atoms using currently available technology.

\end{abstract}


\maketitle

Experiments with ultracold atomic systems have recently realized different incarnations of quantum impurity problems in which a one-dimensional (1D) gas of particles prepared in a particular state (background gas) interacts with a single, distinguishable particle (impurity)~\cite{palzer_impurity_transport_09,bakr_quantum_gas_microscope_2009,weitenberg_spin_addressing_MI_11,
catani_impurity_dynamics_11,fukuhara_spin_impurity_2013,fukuhara_microscopic_2013}.
The background gas exhibits properties that are special to 1D quantum many-body systems~\cite{korepin_book,
gogolin_1dbook,giamarchi_book_1d,cazalilla_one_2011}. Investigations of mobile impurities have contributed to the understanding of various phenomena in those systems, including the excitation spectrum and effective mass~\cite{mcguire_impurity_fermions_65,mcguire_impurity_fermions_66,fuchs_spin_waves_1D_bose_05,
giraud_impurity_09}, orthogonality catastrophe~\cite{castella_mob_impurity_93,castella_mob_impurity_96},
logarithmic diffusion of Green's functions~\cite{zvonarev_ferrobosons_07,akhanjee_ferrobosons_07},
persistence of threshold singularity in spectral functions~\cite{ogawa_mob_impurity_92,nozieres_recoil_94},
its momentum-dependent power-law scaling~\cite{tsukamoto_mob_impurity_98,zvonarev_ferrobosons_07,
matveev_isospin_bosons_08,kamenev_spinor_bosons_09,zvonarev_BoseHubb_09,zvonarev_Yang_Gaudin_09}, and
response to external confinement~\cite{massel_mobile_impurity_TDMRG_2013,peotta_mobile_impurity_TDMRG_2013}
and to external driving~\cite{castroneto_mob_impurity_96,gangardt_Bloch_09,cai_interaction-induced_2010,
johnson_mobile_impurity_lattice_2011,schecter_mobile_impurity_critical_velocity_2012,
schecter_mobile_impurity_Bloch_long_2012}. 
 
In a recent theoretical work~\cite{mathy_flutter_2012}, a phenomenon called quantum flutter 
was reported for an impurity injected with finite momentum $Q$ into a gas of free fermions 
or a gas of Tonks-Girardeau bosons. It was found that the impurity sheds only a part of its momentum to the background gas, and 
forms a correlated state that no longer decays in time. Furthermore, if $Q$ is of the order 
of or larger than the Fermi momentum $k_\mathrm{F}$, the momentum of the impurity undergoes 
long-lived oscillations.  Quantum flutter was demonstrated by examining the full quantum-mechanical 
evolution of the impurity state, obtained from the exact Bethe Ansatz solution, which exploits 
the integrability of the model. Integrability implies the existence of an extensive number of 
mutually commuting integrals of motion, which strongly constrain the dynamics of a 
system~\cite{korepin_book,sutherland_book,gaudin_book}. This raises the general question to 
what extent qualitative results obtained for a particular integrable model are universal. 
As a general rule, the low-energy dynamics of 1D gapless quantum systems does not differ for 
integrable and non-integrable systems~\cite{gogolin_1dbook,giamarchi_book_1d}. However, emerging 
from the time evolution of a far-from-equilibrium initial state, quantum flutter may be viewed 
as a particular case of quench dynamics in a 1D many-body quantum system. Equilibration after a 
quench could be model-specific and could reveal a vast amount of integrability-specific 
phenomena~\cite{kinoshita_quantum_2006, rigol_thermalization_2009,colome_parametric_11}. Whether 
quantum flutter is an integrability and model-specific phenomenon is an open problem, whose analysis 
is especially desirable in view of potential experiments envisioned along this direction.
\begin{figure}
\centering
\includegraphics[width=0.98\linewidth, clip=true,trim= 0 0 0 0]{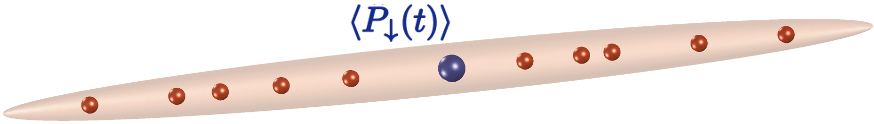}
\caption{(Color online)
Schematic illustration of the system. An impurity atom (large blue sphere) moves with momentum 
$\langle {P}_\downarrow(t) \rangle$ through a background gas of interacting bosonic atoms (small 
red spheres). The shaded area illustrates an equipotential surface confining the atomic motion 
to one dimension. Initially the background gas is prepared in its ground state, and the impurity 
is injected as a plane wave with finite momentum $\langle {P}_{\downarrow} (0) \rangle = Q$. 
The background particles interact with the impurity and with each other through a repulsive 
$\delta$-function potential of strengths $\gamma$ and $\gamma_\mathrm{bg},$ respectively.
}
\label{fig:schematic}
\end{figure}

In this Letter, we report numerical evidence of quantum flutter in the dynamics of an impurity 
with arbitrary mass injected into a 1D quantum gas of interacting bosons, see Fig.~\ref{fig:schematic}. 
The model we use is integrable or non-integrable depending on the choice of parameters. We extract 
the quantum flutter frequency $\omega_\text{f}$ and the saturated impurity momentum 
$\langle {P}_{\downarrow} (\infty) \rangle$ from numerical simulations, for values of impurity mass 
and interaction strength which are accessible in current experiments with ultracold gases. 
We propose an explanation of why $\langle {P}_{\downarrow} (\infty) \rangle$ is non-zero, based on the properties 
of the model at the edge of the excitation spectrum. Moreover, for the integrable case we show that 
$\omega_\text{f}$ is related to the energy difference between two branches of collective excitations of the system.

\textit{Model and numerical method.---}The Hamiltonian of the system schematically illustrated in 
Fig.~\ref{fig:schematic} is
\begin{equation}
H= H_\mathrm{bg} + \frac{ P_\downarrow^2}{2m_\downarrow}+ g \sum_{i=1}^N \delta(x_i-x_\downarrow) \label{eq:Htot}
\end{equation}
where
\begin{equation}
H_\mathrm{bg}= \sum_{i=1}^N \frac{ P_i^2}{2m_\uparrow}+ g_\mathrm{bg} \sum_{1\le i<j\le N} \delta(x_i-x_j). \label{eq:Hbg}
\end{equation}
Here $x_i$ ($ P_i,m_\uparrow$) is the coordinate (momentum, mass) of the $i$th background particle, 
$i=1,\ldots,N,$ and $x_\downarrow (P_\downarrow,m_\downarrow)$ is that of the impurity. Throughout 
this Letter we set $\hbar=1$. We are interested in the limit of large particle number, $N\to\infty,$ 
and system size, $L\to\infty,$ at a fixed background gas density, $\rho_\uparrow=N/L.$ Momenta and 
time are measured in units of Fermi momentum $k_\mathrm{F}$ and Fermi time $t_\mathrm{F}$, respectively:
\begin{equation}
k_\mathrm{F} = \pi \rho_\uparrow, \qquad t_\mathrm{F}= \frac{2m_\uparrow}{k_\mathrm{F}^2}. \label{eq:kFtF}
\end{equation}
The dimensionless strength of the impurity-background repulsion is $\gamma = m_\uparrow g/\rho_\uparrow,$
and background-background repulsion is $ \gamma_\mathrm{bg} = m_\uparrow g_\mathrm{bg}/\rho_\uparrow$.

\begin{figure}
\centering
\includegraphics[width=0.98\linewidth, clip=true,trim= 0 0 0 0]{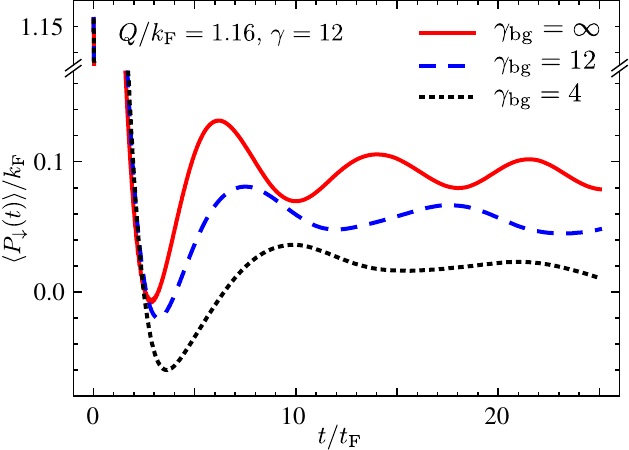}
\caption{(Color online) Impurity momentum $\langle {P}_\downarrow(t) \rangle$ as a function of 
time. Red solid curve: $\gamma_\mathrm{bg}=\infty,$ the integrable Tonks-Girardeau model studied 
in Ref.~\cite{mathy_flutter_2012}. Blue dashed curve: $\gamma_\mathrm{bg}=12,$ the integrable 
bosonic Yang-Gaudin model. Black dotted curve: $\gamma_\mathrm{bg}=4,$ a non-integrable case. 
The initial momentum is $Q = 1.16 k_\mathrm{F}$ and the impurity-background coupling strength 
is $\gamma=12$ for all curves. The masses of the impurity and the background particles are equal, 
$m_\downarrow =m_\uparrow$. All curves exhibit a rapid drop at short times followed by pronounced 
slowly decaying oscillations around a finite saturation value of momentum. We call the frequency 
of these oscillations the quantum flutter frequency $\omega_\text{f}$.}
\label{fig:flutter}
\end{figure}

The impurity is injected into the background gas in a plane wave with momentum $Q$ at time $t=0$, 
so that the initial state of the system is
\begin{equation}
|\mathrm{in}_Q\rangle = c^{\dagger}_{Q\downarrow} |\mathrm{bg}\rangle,
\label{eq:ipw}
\end{equation}
where $|\mathrm{bg}\rangle$ denotes the ground state of the background gas~\eqref{eq:Hbg}. The 
initial state~\eqref{eq:ipw} evolves in time to $|\mathrm{in}_Q(t)\rangle=e^{-itH}|\mathrm{in}_Q\rangle,$ 
where $H$ is the Hamiltonian~\eqref{eq:Htot}. The total momentum of the system, $P_\uparrow+P_\downarrow,$ where $P_\uparrow = \sum_{i=1}^N P_i,$ is conserved. We are interested in the time evolution of the impurity momentum
\begin{equation}
\langle {P}_{\downarrow}(t)\rangle = \langle \mathrm{in}_Q(t)|{P}_{\downarrow}|\mathrm{in}_Q(t) \rangle. \label{eq:Pdown}
\end{equation}
Exemplary plots for integrable and non-integrable cases are shown in Fig.~\ref{fig:flutter}. 
They share the following characteristic of quantum flutter: after a rapid drop pronounced 
slowly decaying oscillations develop, which saturate at a non-zero value of the momentum.

We perform large-scale numerical simulations based on matrix product states (MPS). To this 
end, we finely discretize the Hamiltonian \eqw{eq:Htot} and calculate the initial state 
$|\mathrm{in}_Q\rangle$ with the density matrix renormalization group~\cite{white_density_1992,schollwoeck_density-matrix_2005}. 
The time evolution of the model is then obtained using time-evolving block decimation 
(TEBD)~\cite{vidal_TEBD_2003,vidal_TEBD_2004}. We push TEBD to its limits to perform 
high-accuracy simulations. Specifically, the presented results are obtained for systems 
with $400$ or $600$ sites with $N=40$ or $N=60$ particles and MPS bond dimension 
$M=800$ or $M=600$, respectively. We verified that all of the results are representative 
for the continuum and do not depend on the number of sites, number of particles, or the 
MPS bond dimension.

\begin{figure}
\centering
\includegraphics[width=0.98\linewidth, clip=true,trim= 0 0 0 0]{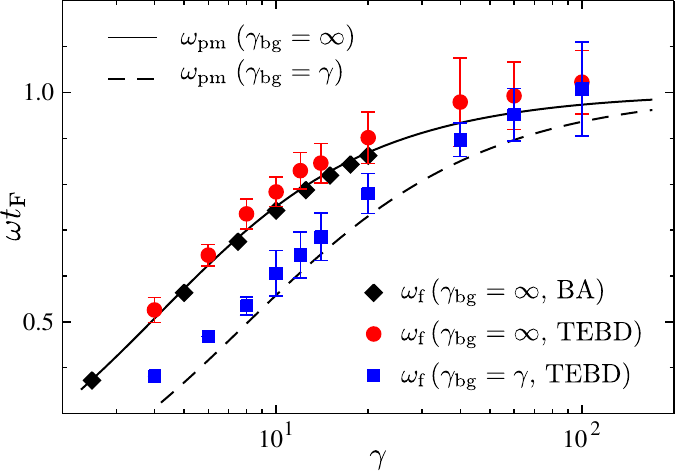}
\caption{(Color online) Quantum flutter frequency for the integrable cases of model~\eqref{eq:Htot}.  Red circles: 
$\gamma_\mathrm{bg}=\infty,$ TEBD simulations. Black diamonds: $\gamma_\mathrm{bg}=\infty,$ 
Bethe Ansatz data from Ref.~\cite{mathy_flutter_2012}. Blue boxes: $\gamma_\mathrm{bg}=\gamma,$ 
TEBD simulations. Each data point and its error bar is obtained by taking twice the 
distance in time between all neighboring extrema of $\langle {P}_\downarrow(t) \rangle$ (exemplary curves of which are shown in Fig.~\ref{fig:flutter}), converting 
them into frequencies, and calculating their mean and standard deviation. Solid (dashed) curve 
is the plasmon-magnon energy difference $\omega_\mathrm{pm}$ for $\gamma_\mathrm{bg}=\infty$ 
($\gamma_\mathrm{bg}=\gamma$) at momentum $k_\mathrm{F},$ obtained from Bethe Ansatz.
}
\label{fig:flutter_frequency}
\end{figure}

\textit{Flutter frequency for integrable cases.---}To elucidate the origin of long-lived 
oscillations in $\langle {P}_\downarrow(t) \rangle$ we compare their periods for two 
integrable cases of model~\eqref{eq:Htot}: Case (a) is the limit of infinite repulsion 
between background particles, $\gamma_\mathrm{bg}=\infty$ (known as a Tonks-Girardeau 
gas~\cite{tonks_complete_1936,girardeau_impurity_TG_60}). It is this integrable case which 
has been used to reveal the quantum flutter phenomenon through Bethe Ansatz and form-factor resummations in 
Ref.~\cite{mathy_flutter_2012}. Case (b) is a particular case of the bosonic Yang-Gaudin model, 
$\gamma_\mathrm{bg}=\gamma$~\cite{gaudin_book,gaudin_fermions_spinful_67,yang_fermions_spinful_67}. 
The data for the oscillation frequency $\omega_\mathrm{f}$ is shown in Fig.~\ref{fig:flutter_frequency}. 
In case (a) we compare $\omega_\mathrm{f}$ obtained from TEBD simulations with the one 
from Bethe Ansatz calculations of Ref.~\cite{mathy_flutter_2012} and find good agreement, 
which is a strong justification of the convergence of the TEBD simulations~\footnote{Comparative 
analyses of TEBD simulations with some alternative high-precision data for an interacting quantum 
many-body system in the continuum are scarce; see also Refs.~\cite{verstraete_continuous_2010,
dolfi_multigrid_2012,stoudenmire_one-dimensional_2012} for studies of ground state properties 
of continuum models.}. In case (b) only data from TEBD is available so far. Our simulations 
demonstrate that oscillations in $\langle {P}_\downarrow(t) \rangle$ develop when $Q$ is of 
the order of or larger than $k_\mathrm{F}$, their amplitude increases with $Q$, and the frequency 
is independent of $Q$.

\begin{figure}
\centering
\includegraphics[width=0.96\linewidth]{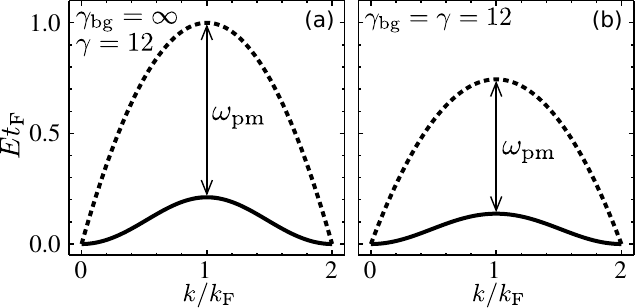}
\caption{Excitation spectrum. Plasmons are the lowest energy 
excitations of the background gas~\eqw{eq:Hbg}. The plasmon dispersion $E_\mathrm{p}(k)$ is shown 
with dotted lines. Magnons are the lowest energy excitations of model~\eqw{eq:Htot}, \ie, background 
gas plus impurity. The magnon dispersion $E_\mathrm{m}(k)$ is shown with solid lines. Two integrable 
cases are illustrated: (a) $\gamma_\mathrm{bg}=\infty$ and (b) $\gamma_\mathrm{bg}=\gamma$. 
The curves are obtained from Bethe Ansatz.}
\label{fig:pm_energy}
\end{figure}

Our interpretation of quantum flutter exploits the structure of the many-body excitation spectra 
of model \eqw{eq:Htot}, which we show in Fig.~\ref{fig:pm_energy}.  The plasmon spectrum is the 
lowest energy excitation of the background gas~\eqw{eq:Hbg} and follows from the Bethe Ansatz 
solution~\cite{lieb_boseI_1963}. The magnon spectrum is the lowest energy excitation of model~\eqref{eq:Htot}
~\footnote{The description in terms of plasmons and magnons used here reduces to the one of 
excitons and polarons used in Ref.~\cite{mathy_flutter_2012} in the limit of infinite repulsion 
between background gas particles.}. For $\gamma_\mathrm{bg}=\infty$ it has been found in 
Ref.~\cite{mcguire_impurity_fermions_65}, and for $\gamma_\mathrm{bg}=\gamma$ it is given explicitly 
in e.g. Refs.~\cite{fuchs_spin_waves_1D_bose_05,zvonarev_Yang_Gaudin_09}. For non-integrable cases 
the magnon spectrum is not yet known, however, techniques proposed in 
Refs.~\cite{porras_renormalization_2006,haegeman_variational_2012} may be used to evaluate it. The 
plasmon-magnon energy difference at the Fermi momentum 
\begin{equation}
\omega_\mathrm{pm}=E_\mathrm{p}(k_\mathrm{F})-E_\mathrm{m}(k_\mathrm{F}) \label{eq:omega_pm}
\end{equation}
is shown in Fig.~\ref{fig:flutter_frequency}. We find that 
\begin{equation}
\omega_\mathrm{f}=\omega_\mathrm{pm} \label{f=pm}
\end{equation}
within numerical accuracy. This striking observation has the following intuitive explanation. 
Provided the impurity is injected in the system with momentum $Q \sim k_\mathrm{F}$, it forms, 
after a few single-particle collisions, a many-body correlated state with the background gas, 
which consists of a superposition of plasmon and magnon excitations at $k_\text{F}$ with nearly 
zero group velocity. The energies of the plasmon and the magnon relative to the zero-momentum 
ground state energy of model~\eqref{eq:Htot} are $E_\mathrm{p}(k_\mathrm{F})$ and $E_\mathrm{m}(k_\mathrm{F}),$ 
respectively. It is precisely the evolution of that correlated state which determines the 
frequency of quantum flutter. If $Q>k_\mathrm{F}$ any momentum in excess of $k_\mathrm{F}$ is 
carried away through an additionally emitted wave packet. If $Q<k_\mathrm{F}$ the aforementioned 
state cannot form, and the oscillations should not develop, which agrees with our numerical observation.

\begin{figure}
\includegraphics[width=0.98\linewidth, clip=true,trim= 0 0 0 0]{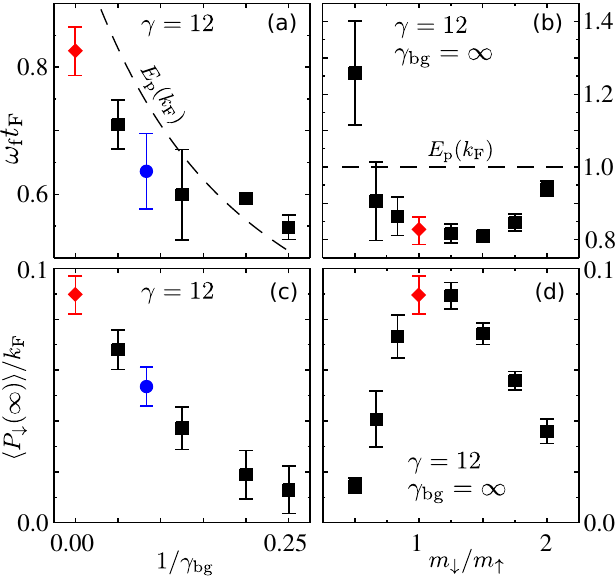}
\centering
\caption{(Color online) Quantum flutter for the non-integrable cases of model~\eqref{eq:Htot}. 
Top panels: flutter frequency $\omega_\mathrm{f}.$ Data is obtained from TEBD simulations 
and error bars are obtained the same way as for Fig.~\ref{fig:flutter_frequency}. The dashed 
line shows the plasmon energy at the Fermi momentum, $E_\mathrm{p}(k_\mathrm{F})$. 
(a) $\omega_\mathrm{f}$ as a function of $1/\gamma_\mathrm{bg}$ for $m_\downarrow=m_\uparrow$ 
and $\gamma=12$. (b) $\omega_\mathrm{f}$ as a function of $m_\downarrow/m_\uparrow$ for 
$\gamma_\mathrm{bg}=\infty$, $\gamma=12$, and $Q=1.16k_\text{F}$. The quantum flutter frequency 
for the integrable Tonks-Girardeau model is indicated by the red diamond and for the integrable 
Yang-Gaudin model by the blue circle. Bottom panels: saturated momentum 
$\langle P_\downarrow(\infty)\rangle$. Data used for panels (c) and (d) are the same as for (a) 
and (b). Error bars indicate the standard deviation of $\langle P_\downarrow(t) \rangle$ after 
the transient decay, $t>15t_\text{F}$.}
\label{fig:flutter_nonintegrable}
\end{figure}

\textit{Flutter frequency for non-integrable cases.---}We investigate $\langle {P}_\downarrow(t) \rangle$ 
when model~\eqref{eq:Htot} deviates from integrability in two different ways: first, $\gamma_\mathrm{bg}$ 
is changed while keeping $\gamma$ constant and second, the mass of the impurity is changed relatively to 
the mass of the background particles. We find that quantum flutter persists in both cases. The flutter 
frequency $\omega_\mathrm{f}$ decreases continuously with decreasing $\gamma_\mathrm{bg}$, 
Fig.~\ref{fig:flutter_nonintegrable}(a). Note that the non-integrable point $\gamma_\mathrm{bg}=20$, 
which lies between the two integrable points $\gamma_\mathrm{bg}=\infty$, red diamond, and 
$\gamma_\mathrm{bg}=\gamma$, blue circle, also follows that trend. One observes $\omega_\text{f} > E_\text{p}(k_\text{F})$ 
for $\gamma_\mathrm{bg}=4$ and $5$, which would imply that $E_\text{m}(k_\text{F})<0$ if one assumes 
that \eq{f=pm} is valid. However, for these background interaction strengths we can only observe 
very few oscillations in $\langle {P}_\downarrow(t) \rangle$ with high enough precision and 
$\omega_\mathrm{f}$ could contain a large systematic error. In the mass-imbalanced case, we find a 
minimum in the flutter frequency as a function of the mass ratio $m_\downarrow/m_\uparrow$, 
\figc{fig:flutter_nonintegrable}{b}. The smallest flutter frequency is obtained for impurities 
that are slightly heavier than the background gas particles. Only very few oscillations in 
$\langle {P}_\downarrow(t) \rangle$ are accessible for $m_\downarrow/m_\uparrow = 0.5,$ which leads 
to the large uncertainty of this data point.

\textit{Saturated momentum.---}We now analyze $\langle {P}_\downarrow(t) \rangle$ in the 
infinite time limit. Bethe Ansatz calculations of Ref.~\cite{mathy_flutter_2012} and TEBD 
simulations reported in this Letter indicate that the amplitude of the oscillations in the 
impurity momentum slowly decays with increasing time. The momentum itself saturates at some 
non-zero value $\langle {P}_\downarrow(\infty) \rangle$ at infinite time, see 
Figs.~\ref{fig:flutter_nonintegrable}(c) and~\ref{fig:flutter_nonintegrable}(d). The physical 
intuition behind the finite value of $\langle P_\downarrow(\infty)\rangle$ can be obtained 
when interpreting the time evolution of the impurity as a sequence of collision events. 
These events create excitations in the background gas which carry away energy and momentum 
of the impurity until it reaches a minimal energy state at some residual momentum, i.e., a 
finite momentum magnon state, indicated by the solid line in \fig{fig:pm_energy}. A consistency 
check of this proposed explanation is that the saturation momentum should be less than the 
maximum possible impurity momentum carried by a magnon state of finite momentum. To this end, 
we examine the impurity momentum in the magnon state $|\mathrm{gs}_q\rangle$ of 
model~\eqref{eq:Htot} with total momentum $q.$ We transform Hamiltonian~\eqref{eq:Htot} 
to the mobile impurity reference frame~\cite{mathy_flutter_2012} and get 
$ H_q= H_\mathrm{bg} + \frac{(q-  P_\uparrow)^2}{2m_\downarrow} + g\sum_{i=1}^N \delta(x_i),$ 
where $H_q= e^{i P_\uparrow x_\downarrow} H e^{-i P_\uparrow x_\downarrow}.$ Writing the magnon energy of the model, $E_\mathrm{m}(q)= \langle\mathrm{gs}_q|H|\mathrm{gs}_q\rangle,$ 
in the mobile impurity reference frame and applying the Hellmann-Feynman theorem we find
\begin{equation}
\langle \mathrm{gs}_q | P_\downarrow| \mathrm{gs}_q\rangle = m_\downarrow v_\mathrm{m}(q), \label{eq:Pgs}
\end{equation}
where
\begin{equation}
v_\mathrm{m}(q) = \left. \frac{\partial E_\mathrm{m}(k)}{\partial k} \right|_{k=q} \label{eq:vmdef}
\end{equation}
is the group velocity of the magnon with momentum $q.$ Equation~\eqref{eq:Pgs} shows that 
the impurity velocity (its momentum divided by its mass) in the magnon state with momentum 
$q$ is equal to the magnon velocity at the same momentum, $v_\mathrm{m}(q)$, which is defined 
solely by the dispersion of the model.

The velocity $v_\mathrm{m}(q)$ is an odd and $2k_\mathrm{F}$-periodic function of $q$ with 
a maximum $v_\mathrm{max}=\max_q v_\mathrm{m}(q)$ at some $q.$ We calculated $v_\mathrm{max}$ 
for the integrable cases of model~\eqref{eq:Htot}, and found 
$v_\mathrm{max} \le k_\mathrm{F}/m_\downarrow $ and that it vanishes as $\gamma\to\infty$ or 
$\gamma_\mathrm{bg}\to 0.$ Comparing it with the estimate of $\langle {P}_\downarrow(\infty) \rangle$ 
from our TEBD simulations we find numerical evidence that
\begin{equation}
\langle {P}_\downarrow(\infty) \rangle  < m_\downarrow v_\mathrm{max}. \label{eq:Pinf}
\end{equation}
For which initial momenta $Q,$ couplings $\gamma$ and $\gamma_\mathrm{bg},$ and mass 
ratio $m_\downarrow/m_\uparrow$ Eq.~\eqref{eq:Pinf} is valid, is an important 
open question. Answering it would clarify the physical intuition that in the 
infinite time limit the impurity velocity is determined by the properties of 
the model near the edge of the excitation spectrum, as is known for 
various other dynamical quantities~\cite{zvonarev_ferrobosons_07,akhanjee_ferrobosons_07,
zvonarev_BoseHubb_09,zvonarev_Yang_Gaudin_09,kamenev_spinor_bosons_09}.

\textit{Summary.---}Our analysis shows strong evidence for the existence of quantum flutter 
away from integrability. The complexity of the TEBD simulations, however, can grow when 
deviating from the integrable points in parameter space~\cite{prosen_is_2007}, which reduces 
the maximum time the simulation is reliable for. Furthermore, close to integrable points the 
dynamics may resemble the integrable one for a long period of time, a phenomenon first 
encountered in the Fermi-Pasta-Ulam problem~\cite{ford_fpu_92}. Quantifying closeness to 
integrability in our model requires a separate study which may help in the understanding of 
effective field theories, as the one suggested for a different setup in Refs.~\cite{gangardt_Bloch_09,
schecter_mobile_impurity_Bloch_long_2012, schecter_mobile_impurity_critical_velocity_2012}. 
Our simulations are ideally suited to model real experimental conditions. For example, the 
setup~\cite{haller_superTonks_2009} consists of about $25$ cesium atoms confined in 1D 
parabolic traps with longitudinal frequency $\sim 2\pi\times 15\mathrm{Hz}$ and highly 
tunable interaction $\gamma_\mathrm{bg}$. We checked that in this case for strong interactions 
about $5$ oscillation periods of $\langle  P_\downarrow(t)\rangle$ should be observable on 
experimentally accessible time scales.

We are grateful to E.~Bogomolny, E.~Burovski, V. Cheianov, E.~Haller, A.~Kamenev, A.~Lamacraft, 
H.C. N{\"a}gerl, and M.~Schecter for fruitful discussions. The authors acknowledge support 
from Harvard-MIT CUA, the DARPA OLE program, AFOSR MURI on Ultracold Molecules, ARO-MURI on 
Atomtronics, the Austrian Marshall Plan Foundation, the Austrian Science Fund (FWF) Project 
No. {J3361-N20} and SFB ViCoM (F41), as well as the Swiss National Science Foundation 
Project No. PA00P2\_126228. Numerical calculations have been performed on the Vienna 
Scientific Cluster.


%

\end{document}